\documentclass[aps,pra,twocolumn,nobalancelastpage]{revtex4-1}
\usepackage[latin9]{inputenc}
\usepackage{color}
\usepackage{bm}
\usepackage{amsmath}
\usepackage{amssymb}
\usepackage{graphicx}
\usepackage{esint}
\usepackage[unicode=true,pdfusetitle,
 bookmarks=false,
 breaklinks=true,pdfborder={0 0 0},backref=false,colorlinks=true]
 {hyperref}
\hypersetup{
 linkcolor=blue, urlcolor=blue, citecolor=blue}
\begin{document}

\title{Engineering topological materials in microwave cavity arrays}

\author{Brandon M. Anderson, Ruichao Ma, Clai Owens, David I. Schuster, Jonathan
Simon}

\affiliation{James Franck Institute and University of Chicago}
\begin{abstract}
We present a scalable architecture for the exploration of interacting
topological phases of photons in arrays of microwave cavities, using
established techniques from cavity and circuit quantum electrodynamics.
A time-reversal symmetry breaking (non-reciprocal) flux is induced
by coupling the microwave cavities to ferrites, allowing for the production
of a variety of topological band structures including the $\alpha=1/4$
Hofstadter model. Effective photon-photon interactions are included
by coupling the cavities to superconducting qubits, and are sufficient
to produce a $\nu=1/2$ bosonic Laughlin puddle. We demonstrate by
exact diagonalization that this architecture is robust to experimentally
achievable levels of disorder. These advances provide an exciting
opportunity to employ the quantum circuit toolkit for the exploration
of strongly interacting topological materials.
\end{abstract}
\maketitle

\section{Introduction}

Despite significant interest in topological physics, the experimental
success in realizing \emph{strongly correlated} topological systems
has thus far been limited to the fractional-quantum hall (FQH) effect
\cite{RevModPhys.71.S298,Stern2008204} in two-dimensional electron
gases. There has recently been significant theoretical explorations
of alternative realizations of FQH systems (and related fractional
Chern insulators \cite{BERGHOLTZ:2013aa,PhysRevX.1.021014}) in the
field of ``quantum engineering.'' Here a topologically nontrivial
Hamiltonian is built from the ground up, typically following a two-ingredient
recipe \cite{BERGHOLTZ:2013aa,PhysRevX.1.021014}: (1) strong interactions
are added to a (2) topologically non-trivial single particle band
structure. This recipe is well motivated theoretically, and experimentally
there has been success in individual implementation of each ingredient.
For example, ultracold atomic systems have experimentally succeeded
in the study of both topologically novel band structures \cite{PhysRevLett.111.185301,PhysRevLett.111.185302,0034-4885-77-12-126401,Jotzu:2014aa,doi:10.1080/00018730802564122},
and strongly interacting but topologically trivial systems \cite{Greiner:2002xy,RevModPhys.80.885}.
However, the simultaneous implementation of both elements, as is necessary
for creating strongly correlated topological states, has thus far
remained elusive.

Photons are a newly emerging platform for quantum engineering. Photonic
crystals have successfully simulated non-interacting band structures
in the regime of RF \cite{PhysRevX.5.021031}, microwave \cite{Wang:2009aa},
or optical \cite{Rechtsman:2013aa,HafeziM.:2013aa} domains. In all
cases, the necessary addition of strong interactions to produce strongly
correlated states has remained challenging. On the other hand, strong
photon-photon interactions are readily achieved in circuit QED experiments
where the exquisite control over few qubit states has allowed the
quantum simulation of, e.g., molecular energies \cite{Martinis2015-Molecule}.
Superconducting circuits have also been advanced as a route to strongly
correlated photonic lattices \cite{Houck:2012aa,PhysRevA.82.043811}.
Implementing a time-reversal-symmetry breaking (TRSB) single particle
band stricture is still necessary to advance towards FQH physics;
current proposals require site-dependent parametric modulation \cite{Fang:2012aa,PhysRevA.82.043811},
which remains experimentally difficult to scale to larger system
sizes. Breaking time reversal symmetry with passive circuit elements
would therefore be a significant advance in engineering of synthetic
quantum materials.

Here we propose a new photonic platform to engineer two-dimensional
tight binding models with non-trival band topology, using arrays of
three-dimensional microwave cavities. Such cavities can be easily
machined from metal to work in the few-to-tens of GHz regime, and
have be shown to provide exceptional coherence at cryogenic temperatures
with quality factors exceeding $Q>10^{8}$ \cite{Reagor2013-highQ}.
The cavities can be tunnel-coupled evanescently by, e.g., directly
milling a channel between two cavities, or using capacitively coupled
transmission lines. Modern machining techniques allow the creation
of scalable lattices with low disorder in on-site energies and tunneling
energies. 

A key component of our scheme is a new technique used to induce the
requisite TRSB flux: this flux is induced by the onsite mode structure
\cite{PhysRevLett.101.246809} of the cavity, in contrast to previous
schemes \cite{HafeziM.:2013aa,Houck:2012aa,Tzuang:2014aa,PhysRevX.5.021031}
that induce a Peierl's phase in the tunnel coupler. The complex amplitude
of the tunnel coupler depends on the local electric field at the periphery
of the cavity. A non-reciprocal (TRSB) phase will require an electric
field with on-site angular momentum, which is coupled to the polarization
of the magnetic field through Faraday's law \cite{Maxwell01011865}.
Therefore, by coupling the magnetic field to a magnetic dipole, say
through use of a ferrimagnetic crystal with a biased magnetic field
\cite{2015arXiv150805290T,PhysRevApplied.2.054002}, a cavity mode
with definite angular momentum can be energetically isolated. In this
way, the ferrite transfers the TRSB bias field to a TRSB flux of the
photonic lattice. This technique is experimentally advantageous compared
to previous schemes because: (1) It allows for the \emph{passive}
creation of topological band structures, and therefore avoids issues
with non-linear mixing with pumping frequency in modulated tunnel
couplings schemes \cite{Fang:2012aa,PhysRevA.82.043811}. (2) The
ferrite couples to a chiral ``bright'' mode, shifting it in energy,
leaving an unshifted chiral ``dark'' mode. The frequency of the
dark mode is first-order insensitive to both the ferrite coupling
strength, loss, and detuning from cavity resonance \cite{2015arXiv150500167L};
it is thus ideal for engineering low-disorder time-reversal breaking
lattice models.

The final ingredient necessary for the study of FQH-like physics is
strong photon-photon interactions. These can be incorporated by coupling
the microwave cavities to superconducting qubits \cite{Houck:2012aa,Reagor-QuantumMemory}.
We consider adding such qubits to each site of the square Hofstdater
model constructed from the linear circuit elements. The Hamiltonian
describing the effective model can be simulated using exact diagonalization
techniques. We therefore numerically explore our system at finite
size and for few photons. The numerical results demonstrate our architecture
will have a FQH eigenstate corresponding to a bosonic $\nu=1/2$ Laughlin
state on a lattice \cite{PhysRevA.76.023613,doi:10.1080/00018730802564122}.

In remainder of this paper we present a scalable architecture for
study of strongly correlated topological materials. The organization
of the paper is as follows: In Sec. \ref{sec:Single-particle-building-blocks}
we discuss the general structure of the microwave cavities needed
for our architecture. We then introduce on-site coupling elements
(Sec. \ref{sub:On-site-symmetry-breaking}) which isolate a desired
eigenmode on each lattice site. When these eigenmodes are tunnel coupled
(Sec. \ref{sub:Tunnel-coupling-and}) an effective flux emerges when
the system is probed in a certain frequency range. This effective
flux can be understood as an emergent gauge field (Sec. \ref{sub:Band-projection-and})
arising from band projection. Concluding the discussion of the general
non-interacting circuit elements, we show this architecture is sufficient
to simulate the Hofstadter model (Sec. \ref{sub:Simulating-a-Hofstadter}),
and present proposals for some other interesting lattice models (Sec.
\ref{sub:Other-lattice-models}.) In Sec. \ref{sec:Exploring-Fractional-Quantum}
we then consider the effects of adding photon-photon interactions
to each site for the purpose of producing fractional Chern insulating
states in finite-size systems. We use a numerical exact diagonalization
technique to simulate our system, and find that the properties of
a $\nu=1/2$ bosonic ``Laughlin puddle'' emerge. Finally, in Sec.
\ref{sub:Robustness-to-disorder} we consider the likely forms of
disorder in the effective tight-binding model description of our system.
We find that the system is insensitive to the largest sources of disorder,
whereas the most sensitive forms of disorder can be controlled. We
conclude that current experimental techniques should be sufficient
to simulate strongly correlated topological systems.

\section{Single-particle building blocks\label{sec:Single-particle-building-blocks}}

\begin{figure}
\includegraphics[clip]{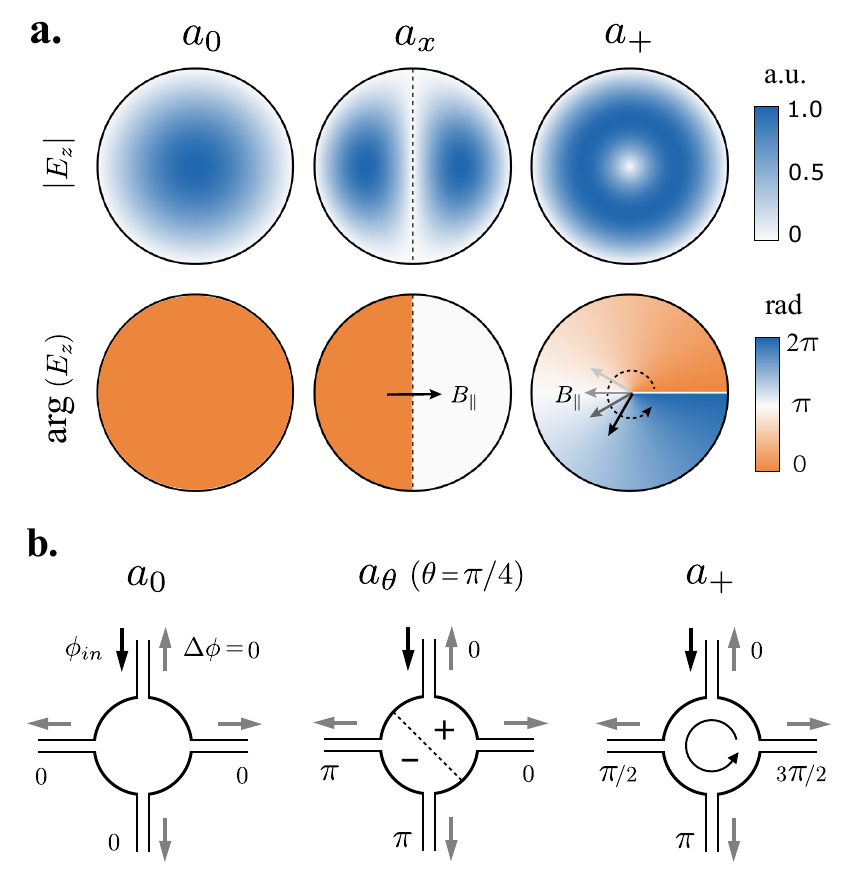}\caption{\label{fig:CavityModes} (a) Structure of the electric field amplitude
and phase (with corresponding magnetic field polarization) of the
modes being considered. The fundamental mode has uniform phase everywhere
with no nodes in the sole electric field component, $E_{z}$. The
two-fold degenerate first excited manifold may be spanned by two modes
with geometrically orthogonal linear-nodes in the $E_{z}$-field ($a_{x}$
and $a_{y}$), or equivalently by two modes exhibiting a single point-node
and $\pm2\pi$ phase winding around the periphery ($a_{\pm})$. For
the modes $a_{x}$ and $a_{+}$, the polarization of the magnetic
field at the cavity center is presented. The polarization of the magnetic
field points along the gradient of the phase of $E_{z}$. For linear
modes the magnetic field is oriented perpendicular to the nodal line,
with direction oscillating in time. For the chiral modes, this magnetic
field polarization precesses in time with frequency $\omega_{0}$.
The size of the circles does not reflect the physical dimension of
the cavities, which is chosen such that the desired mode is always
at frequency $\omega_{0}$. (b) The geometric phase acquired, for
a photon of frequency $\omega_{0}$, when tunneling through a cavity
in which each of type of mode has been isolated. (left) A fundamental
mode cavity will induce no tunneling phase shift between any two contacts.
(center) An isolated  linear mode (via a diagonal conductor or dielectric)
will induce a tunneling phase shift of either $0$ or $\pi$. An isolated
circular mode (via a ferrite in a magnetic field) will produce a phase
shift equal to the relative angle between the incoming and outgoing
tunnel-contacts. }
\end{figure}

We now describe the single particle building blocks for our microwave
architecture. Our goal is to engineer an effective tight-binding Hamiltonian,
$\mathcal{H_{{\rm eff}}}$, for photons whose energy is near $\hbar\omega_{0}$.
We will use a unique cavity eigenmode at this characteristic energy
to represent the tight-binding degree of freedom (and thus restrict
our work to ``spinless'' models.) To model $\mathcal{H_{{\rm eff}}}$
we therefore need an isolated eigenmode on every site, whose frequency
is $\omega_{0}$, with all other onsite eigenmodes far detuned energetically. 

There is significant freedom in the cavity mode structure. We describe
a particular implementation employing transverse magnetic (TM) modes
of cylindrical cavities with non-zero transverse magnetic field $\left(B_{x},B_{y}\neq0,B_{z}=0\right)$
and longitudinal electric field $\left(E_{x},E_{y}=0,E_{z}\neq0\right)$.
We first consider a fundamental cavity mode tuned to frequency $\omega_{0}$
(annihilated by $a_{0}$). The electric field of the fundamental mode,
shown in Fig. \ref{fig:CavityModes}(a), is nodeless and has a spatially
uniform phase across the cavity. This is true regardless of the cavity
geometry. We also consider a \emph{different cavity} where the two-fold
degenerate set of first excited modes (for example, ${\rm TM}_{210}$,
${\rm TM}_{120}$ in a Cartesian basis\cite{pozar2011microwave})
is tuned to a frequency $\omega_{0}$. We define the annihilation
operators of this manifold as $a_{x}$($a_{y}$), according to a node
of the electric field along the $\hat{y}$($\hat{x}$)-axis (see Fig.\ref{fig:CavityModes}(a)).
 Using this notation, the onsite Hamiltonian is $\mathcal{H}^{\left(0\right)}=\omega_{0}a_{0}^{\dagger}a_{0}$,
for a fundamental mode cavity, and $\mathcal{H}^{\left(1\right)}=\omega_{0}\left(a_{x}^{\dagger}a_{x}+a_{y}^{\dagger}a_{y}\right)$
for the excited cavity manifold. Here, and in what follows, we drop
any constants resulting from zero point motion of the electromagnetic
field, and set $\hbar=1$.

While we have written the Hamiltonian in terms of $\hat{x}$/$\hat{y}$
modes, this choice is arbitrary. Analogous to linearly polarized light,
we may rotate the basis by an angle $\theta$: $a_{\theta}=a_{x}\cos\theta+a_{y}\sin\theta$,
and $a_{\theta+\pi/2}=-a_{x}\sin\theta+a_{y}\cos\theta$. Alternatively,
we can construct a basis analogous to circularly polarized light:
$a_{\pm}=\left(a_{x}\pm ia_{y}\right)/\sqrt{2}$, where the phase
changes continuously by $2\pi$ going counter-clockwise (clockwise)
around the cavity center in the $a_{+}$($a_{-}$) mode (see Fig.
\ref{fig:CavityModes}(a)). In this basis, the magnetic field has
an amplitude maximum at the center of the cavity, and a polarization
that lies in-plane and rotates uniformly in time at frequency $\omega_{0}$.

\begin{figure}
\includegraphics[clip]{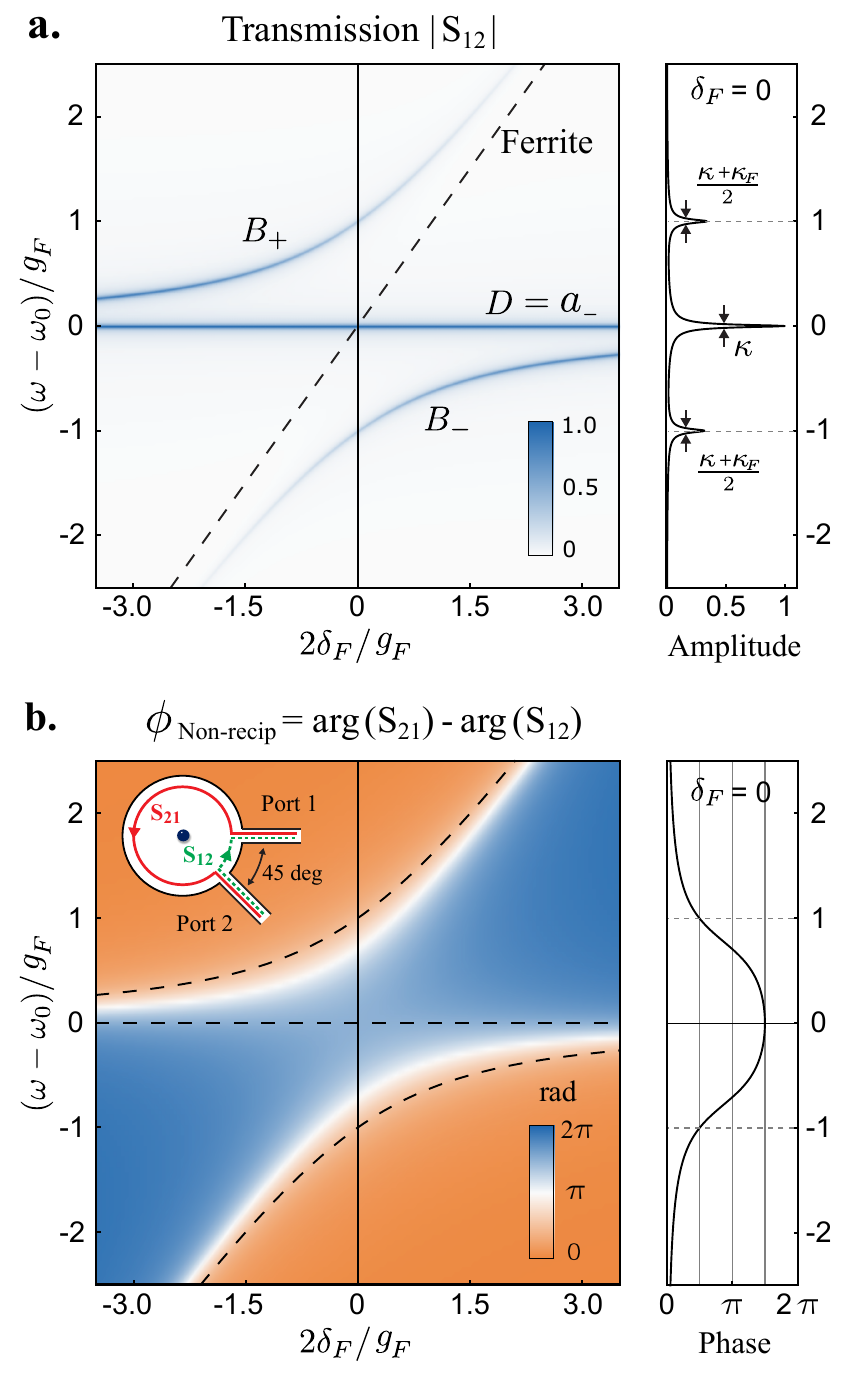}\caption{\label{fig:YIGresponse}Response of a ferrite-cavity site described
by $\mathcal{H}^{\left(F\right)}$ in a pump-probe experiment. Here,
$S_{ij}$ is the response between a port $i=1$ located at the $\hat{x}$
axis, and a second port $j=2$ measured 45 degrees away (inset to
(b)). In order to account for loss, we introduce a linewidth $\kappa$
and $\kappa_{F}$ for the microwave cavity and ferrite respectively.
For physically relevant values of $\kappa/g_{F},\kappa_{F}/g_{F}<10^{-2}$,
this gives a corresponding cooperativity $\eta=4g_{F}^{2}/\kappa\kappa_{F}>10^{4}$.
(a) Amplitude of the transmission response $\left|S_{12}\right|$.
The chirality of a ferrite in a magnetic field allows it to couple
only to the resonator mode of the same handedness, producing an avoided
crossing in the transmission spectrum as the ferrite is tuned through
resonance with the resonator modes. This avoided crossing appears
as two modes which are bright to the ferrite, $B_{+}$ and $B_{-}$.
The dark mode, labeled $D$, does not couple to the ferrite, and is
therefore unshifted and insensitive to ferrite loss $\kappa_{F}$.
(b) The non-reciprocal phase, $\phi_{{\rm Non-Recip}}={\rm arg}\left(S_{21}\right)-{\rm arg}\left(S_{12}\right)$,
reflects the phase difference between going in the 0 degree port and
out the 45 degree port, and the reverse process. This is ideally $\phi_{{\rm Non-Recip}}=3\pi/2$
for the dark mode. This time-reversal break is robust to detuning
in either the ferrite or the probe, and also loss, with the lowest
order correction entering as $\phi_{{\rm Non-Recip}}\approx\frac{3\pi}{2}-2\frac{\delta_{p}(\delta_{p}-2\delta_{F})}{g_{F}^{2}}-\frac{2}{\eta}$.
(c) A slice through the non-reciprocal phase as a function of the
probe detuning, for the ferrite tuned to resonance. The non-reciprocal
phase of $\phi_{{\rm Non-Recip}}\approx3\pi/2$ is apparent near the
dark mode. }
\end{figure}

\subsection{On-site symmetry breaking\label{sub:On-site-symmetry-breaking}}

The on-site degeneracy of $\mathcal{H}^{\left(1\right)}$ can be broken
in a time-reversal preserving or time-reversal breaking manner. The
first type takes the form of violations of cylindrical symmetry. For
example, by placing a thin barrier (e.g. a rectangular conductor or
dielectric) along the nodal line of the $a_{x}$ mode, one adds a
perturbation $\mathcal{V}^{\left(1\right)}=\Delta_{{\rm lin}}a_{y}^{\dagger}a_{y}$
to the Hamiltonian $\mathcal{H}^{\left(1\right)}$. The $a_{x}$ mode
remains (nearly) unperturbed at frequency $\omega_{0}$, while the
orthogonal $a_{y}$ mode will be shifted to higher energy (by an amount
$\Delta_{{\rm lin}}$). This energetically isolates the $a_{x}$ mode
at an energy $\omega_{0}$, and uniquely defines the relative phase
between any two points on the edge of the cavity. (An analogous procedure
isolates $a_{\theta}$ from $a_{\theta+\pi/2}$.) This perturbation
only produces a relative phase of $0$ or $\pi$ between points on
a cavity edge, and therefore cannot break time-reversal symmetry.
Manufacturing imprecision gives rise to perturbations of this type.

For the study of quantum Hall physics, it is necessary to break time-reversal
symmetry; this does not occur naturally for light. We consider achieving
this by employing the coherent magnon interaction between cavity photons
and the spins of a ferrimagnetic material \cite{2015arXiv150805290T,PhysRevApplied.2.054002}:
a small ferrite sphere placed at the center of the cavity with a DC
bias field $\mathbf{B}=+B_{0}\hat{z}$. The ferrite acts as a collective
spin which couples to the magnetic field of the cavity mode and precesses
at a frequency $\omega_{F}=\mu_{0}B_{0}$. (If the bias field is reverse
the following analysis is valid with $a_{+}\leftrightarrow a_{-}$.)
 The magnetic field of the $a_{\pm}$ modes has a maximum at the cavity
center (at the node of the electric field), and an in-plane polarization
that precesses about $\pm\hat{z}$ at a frequency $\omega_{0}$. When
the spin precession frequency $\omega_{F}$ is tuned near $\omega_{0}$,
the polarization of the magnetic field of the $a_{\pm}$ mode rotates
synchronously with the collective spin. The magnetic dipole interaction
results in a strong coupling between the $a_{+}$ mode and the ferrite
mode (denoted $a_{F}$). The magnetic field of the $a_{-}$ mode,
on the other hand, rotates against the collective spin, and thus does
not couple. This results in strong hibridization of the $a_{+}$ mode
with the ferrite, producing two bright magnon modes that are pushed
away from the dark $a_{-}$ mode at frequency $\omega_{0}$. Thus,
when the system is probed near $\omega_{0}$, the isolated dark mode
$a_{-}$ sets a unique magnetic field polarization vector, or equivalently
a unique quantum mechanical phase at the cavity periphery.

The coupling of the cavity-ferrite system is described by the Hamiltonian
$\mathcal{H}^{\left(F\right)}=\mathcal{H}^{\left(1\right)}+\omega_{F}a_{F}^{\dagger}a_{F}+g_{F}\left(a_{+}^{\dagger}a_{F}+a_{F}^{\dagger}a_{+}\right)$,
where $g_{F}$ is the coupling of the ferrite to the $a_{+}$ mode
(and must overwhelm $\Delta_{{\rm lin}}$). A more suggestive notation
is: $\mathcal{H}^{\left(F\right)}=\left(\bm{a}^{\left(F\right)}\right)^{\dagger}H^{\left(F\right)}\bm{a}^{\left(F\right)}$,
where we have defined a ferrite-cavity mode vector $\bm{a}^{\left(F\right)}=\left(a_{x},a_{y},a_{F}\right)^{T}$
and described the cavity-ferrite system with a coupling matrix:
\begin{equation}
H^{\left(F\right)}=\omega_{0}\hat{I}+\begin{pmatrix}0 & 0 & g_{F}/\sqrt{2}\\
0 & 0 & ig_{F}/\sqrt{2}\\
g_{F}/\sqrt{2} & -ig_{F}/\sqrt{2} & 2\delta_{F}
\end{pmatrix},\label{eq:H-F}
\end{equation}
where $2\delta_{F}$ is the detuning of the ferrite from cavity resonance
($\omega_{F}=\omega_{0}+2\delta_{F}$) and $\hat{I}$ is the identity
matrix in the cavity-ferrite space. This Hamiltonian has an uncoupled
dark eigenmode $a_{-}$ that remains at frequency $\omega_{0}$. There
are also two hybridized magnon modes $B_{\pm}=\left(a_{F}\pm a_{+}\right)/\sqrt{2}$
that are frequency shifted by $\omega_{B_{\pm}}-\omega_{0}=\delta_{F}\pm\sqrt{\delta_{F}^{2}+g_{F}^{2}}$.
We henceforth assume the experimentally relevant limit where $g_{F}$
is large compared to $\delta_{F}$, along with any energy scales appearing
in $\mathcal{H}_{{\rm eff}}$.

Key features of this system are understood by examining the cavity
response using input-output formalism \cite{scully1997quantum,2015arXiv150805290T}:
We consider response at a small detuning $\delta_{p}=\omega-\omega_{0}$,
and ferrite detuning $2\delta_{F}$, from $\omega_{0}$, and introduce
a finite lifetime of the cavity modes $\left(\kappa\right)$ and ferrite
mode $\left(\kappa_{F}\right)$, with a resulting cooperativity $\eta=4g_{F}^{2}/\kappa\kappa_{F}$.
The amplitude of the response is shown in Fig. \ref{fig:YIGresponse}(a).
Here, there are three maxima in the response as $\delta_{p}$ is varied,
corresponding to three eigenmodes: the $a_{-}$ mode remains at $\omega_{0}$
for all ferrite detunings, while the $a_{+}$ mode mixes with the
ferrite mode $a_{F}$ and undergoes an avoided crossing as the ferrite
detuning is swept. When probed at frequencies near $\omega_{B_{\pm}}$,
the bright magnon modes have the TRSB phase winding. Importantly,
though only the bright mode couples to the ferrite, both the bright
and dark modes are non-reciprocal. In fact the dark mode is preferred
because crucially, it is insensitive to loss in the ferrite and is
less sensitive to detuning and linear disorder. When the ferrite is
exactly on resonance ($\delta_{F}=0$), the dark mode has an exact
TRSB phase winding of $2\pi$ at the periphery of the cavity. At probe
frequencies different than $\omega_{0}$ the phase winding ceases
to be uniform. However, this nonuniformity is quadratically insensitive
in both the ferrite detuning and the energy of the probe photon. This
can be seen from the saddle-point structure in Fig. \ref{fig:YIGresponse}(b).
Here the non-reciprocal phase response of this system is calculated
between two cavity contacts separated by an angle of $\pi/4$ on the
cavity periphery (inset); the difference between forward and backwards
phases is $\phi_{{\rm Non-Recip}}\approx\frac{3\pi}{2}-2\frac{\delta_{p}(\delta_{p}-2\delta_{F})}{g_{F}^{2}}-\frac{2}{\eta}$.
This quadratic insensitivity of the eigenmode structure significantly
decreases disorder in $\mathcal{H}_{{\rm eff}}$ arising from disorder
in the ferrite, cavity, or bias field. 

We now define our onsite Hamiltonian at a site $j$ with the notation
$\mathcal{H}_{{\rm os},j}^{\left(\alpha\right)}=\left(\bm{a}_{j}^{\left(\alpha\right)}\right)^{\dagger}H_{j}^{\left(\alpha\right)}\bm{a}_{j}^{\left(\alpha\right)}$
where $\alpha=0$ corresponds to a fundamental cavity, $\alpha=\theta$
defines a site isolating a linear mode $a_{\theta}$, and $\alpha=F$
corresponds to a ferrite site. The form of the vector of annihilation
operators on site $j$ depends on the type of site with $\bm{a}_{j}^{\left(0\right)}=a_{j0}$,
$\bm{a}_{j}^{\left(\theta\right)}=\left(a_{jx},a_{jy}\right)^{T}$,
and $\bm{a}_{j}^{\left(F\right)}=\left(a_{jx},a_{jy},a_{jF}\right)^{T}$
for a fundamental, linear, and ferrite site respectively. The onsite
coupling Hamiltonians are $H_{j}^{\left(0\right)}=\omega_{0}$, 
\[
H_{j}^{\left(\theta\right)}=\begin{pmatrix}\omega_{0} & -\Delta_{{\rm lin}}e^{i\theta}\\
-\Delta_{{\rm lin}}e^{-i\theta} & \omega_{0}
\end{pmatrix},
\]
and $H_{j}^{\left(F\right)}$ is given by Eq. (\ref{eq:H-F}). A system
of uncoupled cavities is then generically described by the on-site
Hamiltonian $\mathcal{H}_{{\rm os}}=\sum_{j}\mathcal{H}_{{\rm os,}j}^{\left(\alpha\right)}=\sum_{j}\left(\bm{a}_{j}^{\left(\alpha\right)}\right)^{\dagger}H_{j}^{\left(\alpha\right)}\bm{a}_{j}^{\left(\alpha\right)}$,
where each $\alpha_{j}$ represents one of the distinct cavity types.

\subsection{\textit{\emph{Tunnel coupling and nontrivial flux\label{sub:Tunnel-coupling-and}}}}

Coupling of the cavities is realized by connecting their edges with
evanescent waveguides. The amplitude of such a tunnel coupling is
determined by the geometry (length and cross-section) of the channel.
We emphasize that the coupler is only virtually populated -- the evanescent
wave ``propagates'' with a purely imaginary wavevector -- so the
phase of the photon is spatially uniform throughout the channel, and
the channel has no dynamical degree of freedom. 

Now consider a site with an isolated eigenmode (such as $a_{\theta}$
or $a_{\pm}$) at frequency $\omega_{0}$, that has a nonuniform phase
profile. Attached to this cavity are two or more tunnel contacts to
nearby cavities. The nonuniform phase of the mode implies that a photon
that tunnels in along one channel will acquire a geometrical phase
shift as it tunnels out along a different channel; this is the origin
of the induced TRSB flux. 

In a fundamental mode cavity, both the amplitude and phase around
the edge are uniform, and therefore no geometric phase will be acquired
for two tunneling contacts, regardless of where they are attached.
In contrast, a ferrite-cavity site with two tunneling channels separated
by an angle $\phi$ on the cavity perimeter will experience a geometrical
phase shift of $\pm\phi$ for an isolated $a_{\pm}$ mode (see Fig.
\ref{fig:CavityModes}(b).)

On the other hand, two tunnel contacts on a site with a mode $a_{\theta}$
will experience a relative phase shift only if the photon crosses
a node between the processes of tunneling in and tunneling out (see
Fig. \ref{fig:CavityModes}(b).) For an $a_{\theta}$ mode, the local
amplitude at the cavity periphery will be nonuniform and the tunneling
magnitude will gain position dependence (see Figs. \ref{fig:CavityModes}(a).)
(An ideal point contact at an angle $\phi$ relative to the $\hat{x}$
axis will tunnel with amplitude $t\sim\cos\left(\theta-\phi\right)$
at frequencies near $\omega_{0}$.) This is in contrast to a mode
$a_{0}$ or $a_{\pm}$, for which the mode magnitude, and thus the
tunneling magnitude, is uniform. 

For an open 1D chain of tunnel-coupled cavities, the geometric phase
arising from onsite elements can be eliminated through a gauge transformation.
On the other hand, closing the loop produces a net phase (flux) after
tunneling around a closed loop (plaquette). This flux is analogous
to a discrete version of Berry's phase and cannot be eliminated through
a gauge transformation. Rather it will, appear on the tunnel coupling
terms in $\mathcal{H}_{{\rm eff}}$. In this way, a variety of lattice
models can be realized that have nontrivial Peierl's phases.

\subsection{Band projection and geometric phase\label{sub:Band-projection-and}}

The arguments above suggest there is a nontrivial Berry's flux through
a plaquette when energy scales are restricted to a small window around
$\omega_{0}$ (and onsite perturbations are sufficiently large). At
energies away from $\omega_{0}$ the full degrees of freedom must
be considered. The Hamiltonian that describes this general system
is given by $\mathcal{H}_{0}=\mathcal{H}_{T}+\mathcal{H}_{{\rm os}}$,
where $\mathcal{H}_{T}=\sum_{\left\langle ij\right\rangle }\left(\bm{a}_{i}^{\left(\alpha_{i}\right)}\right)^{\dagger}T_{ij}\bm{a}_{j}^{\left(\alpha_{j}\right)}$
tunnel couples the lattice of cavities defined by $\mathcal{H}_{{\rm os}}$
in the previous section. The matrix $T_{ij}$ represents the tunneling
between two neighboring sites with onsite mode structure $\alpha_{i}$
and $\alpha_{j}$. The specific form depends on the type of tunneling,
but we note that an ideal point contact will have a non-zero overlap
for only a single bare cavity mode (not a coupled eigenmode) at frequency
$\omega_{0}$. For fundamental cavities this is naturally $a_{0}$,
whereas in first excited cavities only one linear combination of $a_{x}$
and $a_{y}$ will have a nonzero contribution to $T_{ij}$. All other
tunneling terms will vanish.

The geometric flux can now be rigorously calculated considering a
unitary $U_{j}^{\left(\alpha\right)}$ matrix at every site $j$ that
locally diagonalized $H_{j}^{\left(\alpha\right)}$: $H_{j}^{\left(\alpha\right)}=U_{j}^{\left(\alpha\right)}\Delta_{j}^{\left(\alpha\right)}\left(U_{j}^{\left(\alpha\right)}\right)^{\dagger}$,
where $\Delta_{j}^{\left(\alpha\right)}$ is a diagonal matrix of
onsite energy eigenvalues with a unique mode at frequency $\omega_{0}$.
Applying this unitary rotation to every site then transforms the Hamiltonian
to $\mathcal{H}_{{\rm os}}=\sum_{j}\left(\tilde{\bm{a}}_{j}^{\left(\alpha_{j}\right)}\right)^{\dagger}\Delta_{j}^{\left(\alpha_{j}\right)}\tilde{\bm{a}}_{j}^{\left(\alpha_{j}\right)}$,
and $\mathcal{H}_{T}=\sum_{\left\langle ij\right\rangle }\left(\tilde{\bm{a}}_{i}^{\left(\alpha_{j}\right)}\right)^{\dagger}\tilde{T}_{ij}\tilde{\bm{a}}_{j}^{\left(\alpha_{j}\right)}$,
where $\tilde{\bm{a}}_{j}^{\left(\alpha_{j}\right)}=U_{j}^{\left(\alpha_{j}\right)}\bm{a}_{j}^{\left(\alpha_{j}\right)}$
and $\tilde{T}_{ij}=U_{i}^{\left(\alpha_{i}\right)}T_{ij}\left(U_{j}^{\left(\alpha_{j}\right)}\right)^{\dagger}$.
This transformation has the effect of locally diagonalizing each onsite
Hamiltonian at the cost of transforming the tunneling matrix between
sites.

The restriction to the unique mode at frequency $\omega_{0}$ will
amount to a projection into the onsite modes $\hat{a}_{j0}$. Applying
a projection operator $\mathcal{P}_{j0}$ to eliminate unoccupied
modes results in $\mathcal{H}_{{\rm eff},0}\equiv\sum_{j}\mathcal{P}_{j0}\mathcal{H}_{0}\mathcal{P}_{j0}=\sum_{j}\omega_{0}\hat{a}_{j0}^{\dagger}\hat{a}_{j0}+\sum_{\left\langle ij\right\rangle }\hat{t}_{ij}\hat{a}_{i0}^{\dagger}\hat{a}_{j0}$,
where the effective tunneling matrix $\hat{t}_{ij}$ can contain a
nontrivial phase. We emphasize that while this tunneling phase is
gauge dependent, the net flux through a plaquette naturally remains
gauge invariant.

We now discuss how the circuit elements described above can result
in interesting band structures for a frequency range around $\omega_{0}$.
For a plaquette consisting of only fundamental mode cavities, a photon
acquires no net phase going around a plaquette (corresponding to zero
flux.) An $a_{\theta}$ mode cavity would contribute a net flux of
$\pi$ to a plaquette if the tunneling contacts cross the nodal line,
and $0$ otherwise.

For plaquettes with ferrite sites, the TRSB flux is directly proportional
to the angle between the two tunneling channels attached to the ferrite
cavity. Thus, a resonator with $N$ evenly spaced tunneling contacts
will contribute a flux of $\phi=2\pi/N$ to each adjacent plaquette
connected by tunnel couplers. This will allow for simulation of $\alpha=1/N$
Hofstadter models in different lattice geometries, such as $\alpha=1/4$
in a square geometry and $\alpha=1/6$ in a triangle. In general,
using only the first excited modes we are limited to a total flux
of $2\pi$ per internal ferrite. However, using higher order modes
allows for a net flux of any integer multiple of $2\pi$ per plaquette. 

The nontrivial effective flux is a discrete geometric (Berry's) phase;
it is a direct lattice analogue to ``synthetic gauge fields'' studied
in ultracold atomic systems \cite{0034-4885-77-12-126401}. There,
Raman fields are used to couple internal (spin) degrees of freedom.
The Raman fields vary slowly in space, but also provide a large energetic
separation between dressed states. After preparing the system in a
single dressed state, the dynamics respond analogously to a system
under the influence of a nontrivial external gauge potential. Similar
to the study of synthetic gauge fields in ultracold atoms, the microwave
lattice scheme can also be extended to the regime of synthetic non-Abelian
gauge fields, such as spin-orbit coupling \cite{PhysRevLett.95.010404,0034-4885-77-12-126401};
we leave such systems to future work.

\begin{figure*}
\includegraphics{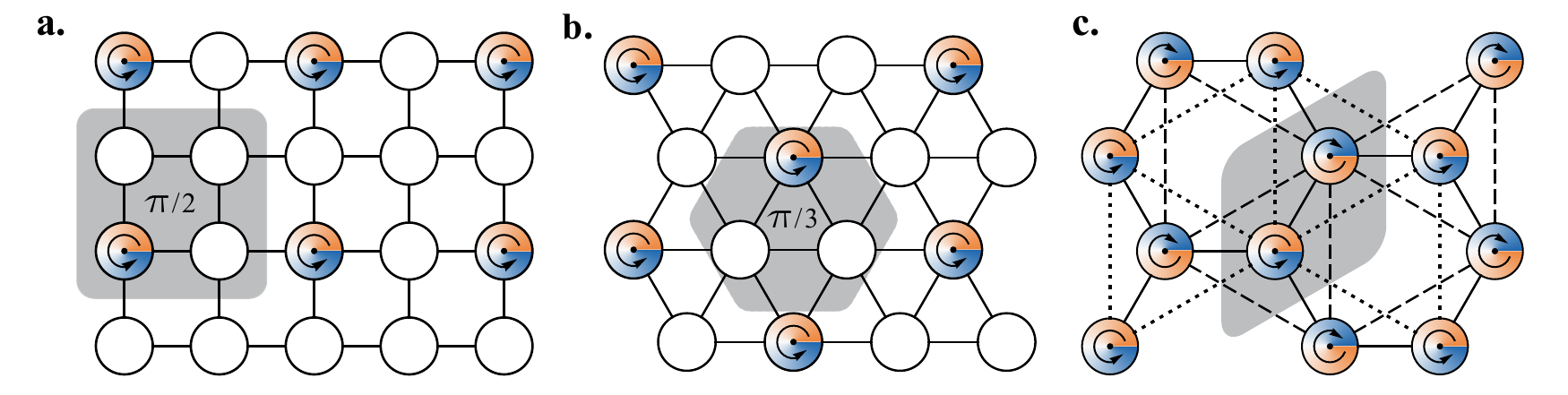}

\caption{Implementations of topological band structures described in the main
text. (a) Hofstadter model with $\alpha=1/4$: a square, four-site,
unit cell has one ferrite and three fundamental resonators. In the
effecitive model near a frequency $\omega_{0}$, the ferrite induces
a quarter flux quantum for each of the four neighboring plaquettes.
(b) Triangular Hofstadter model with $\alpha=1/6$: a triangular lattice
with a three site unit cell has a single ferrite. This ferrite touches
each of six neighboring plaquettes. (c) Haldane model: a hexagonal
lattice has a ferrite on each lattice site. By staggering the handedness
of the dark modes on A and B sublattices, the net flux vanishes. Addition
of next-nearest-neighbor tunneling terms breaks time-reversal symmetry
by inducing a flux of $\alpha=\pm\pi/3$ in each sublattice. \label{fig:Other-Models}}
\end{figure*}

\begin{figure}
\includegraphics{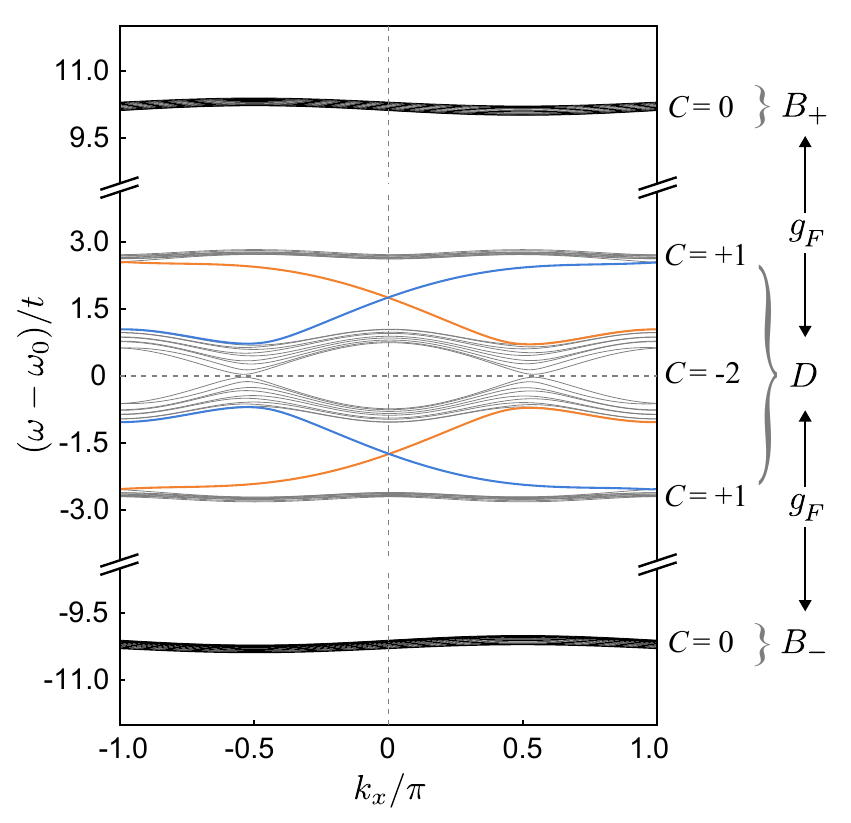}\caption{\label{fig:Hofstadter}Band structure of the microwave cavity implementation
of the $\alpha=1/4$ Hofstdater model in a strip geometry. The unit
cell presented in Fig. \ref{fig:Other-Models}(a) has three fundamental
mode cavities, along with a single ferrite site, resulting in a total
of six bands. When the ferrite energy is tuned large, two bright bands
(black) are far separated from the four bands (gray) bands near $\omega_{0}$.
These four bands have an edge state and Chern number structure consistent
with the $\alpha=1/4$ Hofstadter model. Here, blue (orange) lines
connecting the bulk bands represent edge modes localized on the left
(right) edge. No edge states connect the dark bands to the bright
bands.}
\end{figure}

\subsection{Simulating a Hofstadter model\label{sub:Simulating-a-Hofstadter}}

Given the circuit elements described above, it is now straightforward
to construct an $\alpha=1/4$ Hofstadter model. Of the many equivalent
configurations, we present one that requires the minimum number of
ferrites per unit cell. 

Consider a square lattice of fundamental resonators, with every fourth
fundamental cavity replaced by ferrite cavity, such that all plaquettes
include one ferrite. This results in a square four-site unit cell.
(See shaded area in Fig. \ref{fig:Other-Models}(a).) This is consistent
with the fact that a Hofsdater model with flux $\alpha=p/q$ will
have a $q$-site unit cell. When the system is probed at frequencies
near $\omega_{0}$, the time-reversal symmetry breaking mode will
contribute a phase of $\pi/2$ into each of the four neighboring plaquettes.
Since each plaquette touches only a single ferrite the total flux
will be uniform $\alpha=\left(\pi/2\right)/2\pi=1/4$. 

We have verified this system provides the expected Hofstadter physics
in two ways. First, we have explicitly calculated the band projected
model for the phase convention defined above. The net flux through
a plaquette is found to be $\pi/2$ as expected. Second, we have numerically
diagonalized the Hamiltonian including all relevant degrees of freedom.
We chose the ferrite-cavity coupling to be ten times the tunneling
energy ($g_{F}=10t$), and consider a strip geometry. As seen in Fig.
\ref{fig:Hofstadter}, the full model has six bands. The top and bottom
(black) bands are composed almost entirely of the bright magnon modes.
These bands are only weakly dispersive, as tunneling between bright
modes must occur off-resonantly through neighboring fundamental cavities,
and is therefore suppressed to $\sim t^{2}/g_{F}$.

In contrast, the middle four (gray) bands are composed almost entirely
of dark and fundamental cavity modes. These states form the effective
Hofstadter model, and have all expected properties: The four bands
are energetically symmetric around the energy $\omega_{0}$; the top
and bottom band are gapped relative to the two middle bands, which
touch at two distinct Dirac points; the band Chern numbers for the
continuum model are calculated to be (from top to bottom) $C=0,1,-2,1,0$,
consistent with topologically trivial bright bands sandwiching an
$\alpha=1/4$ Hofstadter model in the dark sector. A finite size calculation
shows chiral edge channels that emerge between the Hofstadter bands,
but not between the dark (Hofstadter) and bright bands.

\subsection{Other lattice models\label{sub:Other-lattice-models}}

These circuit elements are powerful for implementing a variety of
topological band structures. In passing we present two additional
examples. Figure \ref{fig:Other-Models}(b) is an implementation of
a triangular Hofstadter model with $\alpha=1/6$, where we use a triangular
lattice with a single ferrite placed in each three site unit cell.
Using arguments similar to above, this introduces a net flux of $\alpha=1/6$
per plaquette, as desired. 

Tunnel couplings may additionally cross by, e.g., machining channels
on both the top and bottom of the substrate supporting the cavity
array. This allows for implementation of an even larger class of topologically
non-trivial single particle Hamiltonians. In Fig. \ref{fig:Other-Models}(c)
the Haldane model is constructed from a hexagonal lattice with next-neighbor
tunnel couplings added, where a ferrite is added to every site. By
alternating the sign of the DC bias field, and thereby the chirality
of the ferrite dark modes, between A and B sites the net flux per
plaquette vanishes, while time reversal symmetry is locally broken.
This geometry specifically produces the ``ideal'' flux configuration
that is gauge equivalent to a flux quantum per plaquette \cite{PhysRevLett.106.236804}.

In addition to the examples presented here, other frequently studied
Hamiltonians such as the 2D chiral-$\pi$ \cite{PhysRevLett.106.236804},
or 1D chiral models such as the SSH model may also be directly implemented
\cite{PhysRevLett.42.1698}. The circuit architecture described above
is useful for realization of a variety of paradigmatic lattice models
described in the literature. The possibilities are extensive and we
leave elucidation of these to future works.

\section{Exploring Fractional Quantum Hall Physics\label{sec:Exploring-Fractional-Quantum}}

\begin{figure}[!t]
\includegraphics[width=1\columnwidth]{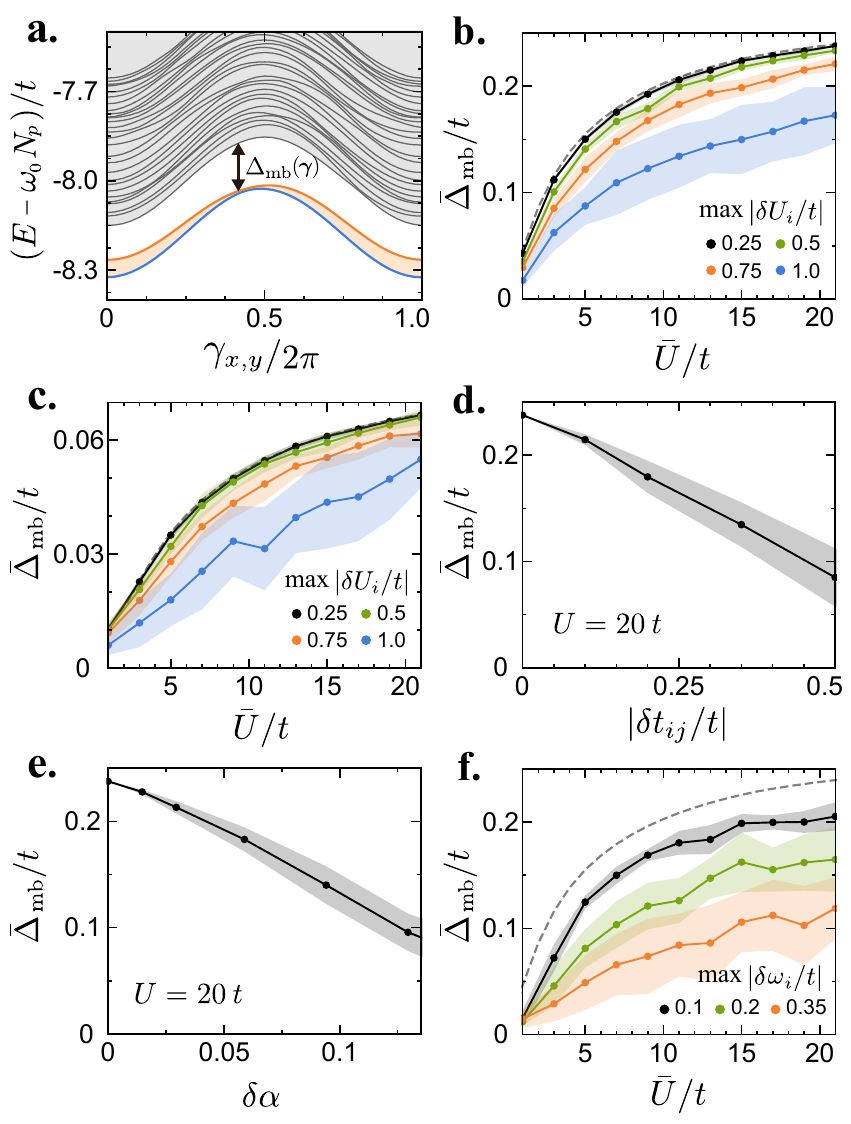}\caption{\label{fig:FQH-numerics}(a) Spectral flow of a of the lowest 30 many-body
eigenstates as a flux $\gamma_{x}=\gamma_{y}=\gamma$, $\gamma\in\left[0,2\pi\right]$
is inserted into $\mathcal{H}_{{\rm eff},{\rm mb}}$. Weak disorder
is included that visually splits the numerically exact two-fold degeneracy
of the ground state manifold in the clean limit. For all flux, the
GSM (orange, blue lines) remains gapped relative to the (shaded gray)
excited states. Provided this gap remains in the presence of disorder,
a many-body Chern number can be defined. The disorder-averaged minimum
value of this gap is denoted $\bar{\Delta}_{{\rm mb}}$. This is used
as a metric to test sensitivity to various types of disorder. (b)
$\bar{\Delta}_{{\rm mb}}$ as a function of the average interaction
strength $\bar{U}$ for various strengths of interaction disorder
$\delta U_{i}$. The many-body gap is not significantly suppressed
at large interaction strengths. (c) Same as (b), but all ferrite sites
are made non-interacting. This comes at the cost a drop of $\bar{\Delta}_{{\rm mb}}$
by a factor of about four. (d) Effect of disorder in the magnitude
of the tunneling strength $\left|\delta t_{ij}\right|$, in a strongly
interacting case ($U_{i}=20$). (e) Effect of disorder in the flux
$\delta\alpha$ through each plaquette. (f) Effect of disorder $\delta\omega_{i}$
in the frequency of the onsite eigenmode near $\omega_{0}$. All error
bars (shaded regions) are standard errors of the mean for about 30
realizations of the disorder; the disorders $\delta U_{i}$, $\left|\delta t_{ij}\right|$,
and $\delta\omega_{i}$ are uniformly distributed, whereas $\delta\alpha$
is the half-width half-max of a Gaussian distribution.}
\end{figure}

The microwave architecture above is linear and should be equally effective
at simulating non-interacting topological structures in both the classical
and few-photon limits. On the other hand, to create strongly correlated
topological states, strong photon-photon interactions are required.
It is well established that a nonlinear interaction between photons
arises through coupling the microwave cavities to Josephson junction
qubits \cite{Wallraff:2004qf,Reagor-QuantumMemory}. The addition
of a qubit to each site of the synthetic $\alpha=1/4$ Hofstadter
model proposed above should be sufficient to explore bosonic fractional-quantum
hall physics \cite{PhysRevA.76.023613,doi:10.1080/00018730802564122}
for microwave photons. We expect that the addition of interactions
should result in a FQH-like eigenstate lying at the edge of the dark
bands, and well separated from the bright bands. 

These FQH could then be explored in a variety of ways. For example,
a pump-probe experiment with low average photon number near the energies
of the effective FQH state will have a small probability of exciting
the FQH manifold. Alternatively, a chemical potential for photons
\cite{hafezi2015chemical} could be used to drive the system to a
steady state with the precise number of photons requisite for a specific
geometry. In finite size systems, the desired FQH eigenstates will
be ones with definite photon number. We note that these preparation
schemes precisely set the photon number, and thus the assumption of
definite photon number below is experimentally relevant.

We now demonstrate the existence of bosonic FQH eigenstates in our
architecture by numerically exploring the combined single-particle
and interacting Hamiltonian $\mathcal{H}_{{\rm eff},{\rm mb}}=\mathcal{H}_{{\rm eff},0}+\mathcal{H}_{{\rm eff},I}$
projected to describe energies near $\omega_{0}$. Provided the strength
of the interactions is weak compared to the ferrite-cavity coupling,
a projected qubit-mediated interaction Hamiltonian can be used: $\mathcal{H}_{{\rm eff},I}=\frac{1}{2}\sum_{i}U_{i}\hat{n}_{i0}\left(\hat{n}_{i0}-1\right)$
where $U_{i}$ is the effective photon-photon interaction for the
photon number $\hat{n}_{i0}$ of the unique mode near $\omega_{0}$
at site $i$. We will work at a fixed particle number relevant for
both numerical calculation and experiments where an exact number of
photons can be prepared. 

We consider a finite-size geometry with $N=N_{x}\times N_{y}$ lattice
sites and toroidal boundary conditions, producing a degenerate ground
state manifold. Such boundary conditions need not be a purely theoretical
construct: They could be implemented explicitly by either machining
the cavities on a physical torus, or connecting the opposite boundaries
with waveguides \cite{ningyuan2015time}. In order to maintain a uniform
effective flux each plaquette must touch exactly one ferrite. This
geometric restriction forces $N/4$ to be an integer, and thus that
the projected model has $N_{\phi}=\alpha N=N/4$ flux quanta. For
a system of fixed particle number $N_{p}$, a FQH ground state is
expected to emerge when the filling factor $\nu=N_{p}/N_{\phi}$ is
expressible as a certain sequence of rational numbers. The most stable
(largest many-body gap) filling is desirable; this occurs at a filling
$\nu=1/2$, which further restricts $N_{p}=N/8$ particles ($N$ divisible
by eight). It is experimentally advantageous to start with few photon
states; we thus explore small geometries with $N_{x}=4$ and $N_{p}=2,3,4$
(and therefore $N_{y}=2N_{p}$); all numerical results presented below
are for $N_{p}=3$ and $N_{y}=6$. (Interestingly, this geometry produces
non-dispersive bands with $N_{y}=4$, which are only weakly dispersive
when $N_{y}=6,8$ \cite{PhysRevB.90.115132}.)

As explored in previous works \cite{PhysRevLett.108.066802,BERGHOLTZ:2013aa,PhysRevB.90.115132,PhysRevX.1.021014},
many properties of the FQH are expected to survive in small few-particle
geometries. To demonstrate this for our system, we start by exactly
diagonalizing $\mathcal{H}_{{\rm eff},{\rm mb}}$ in the clean limit
with $U_{i}=U\gg t$. The geometries described above result in a two-fold
degeneracy (exact to numerical precision) in the many-body ground
state manifold (GSM), as expected for a $\nu=1/2$ Laughlin state
\cite{RevModPhys.71.S298,PhysRevA.76.023613}. We label the states
in the GSM as $\left|\Psi_{1}\right>$ and $\left|\Psi_{2}\right>$;
translational invariance allows these states to be distinguished by
their center-of-mass momentum \cite{PhysRevX.1.021014}. It will be
useful to impose twisted boundary condition: $\Psi_{m}\left(x+N_{x},y\right)=e^{i\gamma_{x}}\Psi_{m}\left(x,y\right)$
and $\Psi_{m}\left(x,y+N_{y}\right)=e^{i\gamma_{y}}\Psi_{m}\left(x,y\right)$
for all eigenstates $\Psi_{m}\left(x,y\right)$, both in the GSM ($m=1,2$)
and not in the GSM ($m\geq3$). The phases $\gamma_{x}\times\gamma_{y}$
represent an Aharanov-Bohm flux $\gamma_{x,y}\in[0,2\pi)$ adiabatically
inserted along the $\hat{x}$, $\hat{y}$ axes of the torus (as could
be implemented with rf-modulated tunneling \cite{Fang:2012aa,Tzuang:2014aa}.) 

Figure \ref{fig:FQH-numerics}(a) shows the spectral flow of the lowest
30 eigenstates as a flux quantum is simultaneously inserted along
both axes of the torus. Throughout this process the GSM remains. (Here,
we include disorder to visually split the two ground states; in the
clean limit the GSM is degenerate to numerical accuracy for all flux
values.) For each flux $\bm{\gamma}$ we define the spectral gap to
the first set of excited states: 
\begin{equation}
\Delta_{{\rm mb}}\left(\bm{\gamma}\right)=\min_{m\notin{\rm GSM}}E_{m}\left(\bm{\gamma}\right)-\max_{m\in{\rm GSM}}E_{m}\left(\bm{\gamma}\right),
\end{equation}
and we find that the spectral gap is approximately constant for all
flux $\bm{\gamma}=\left(\gamma_{x},\gamma_{y}\right)\in{\rm MBZ}$,
where ${\rm MBZ}$ is the many-body Brillouin zone. Since the GSM
remains gapped through the full flux insertion process, we can define
the many-body Chern number \cite{PhysRevB.31.3372}:
\begin{equation}
C_{m}=\frac{1}{4\pi}\int_{\bm{\gamma}\in{\rm MBZ}}d^{2}\bm{\bm{\gamma}}\,\left[\nabla_{\bm{\bm{\gamma}}}\times\left\langle \Psi_{m}\left(\bm{\bm{\gamma}}\right)\left|\nabla_{\bm{\bm{\gamma}}}\right|\Psi_{m}\left(\bm{\bm{\gamma}}\right)\right\rangle \right],
\end{equation}
for a state $m\in{\rm GSM}$. This Chern number will evaluate to an
integer and give a quantized hall conductance. For a clean system,
both states in the GSM can be identified by their center-of-mass momentum
and $C_{m}$ calculated independently. In this case we find they each
carry a fractionalized many-body Chern number of $C=1/2$, consistent
with the thermodynamic limit. This result is also consistent with
prior numerical studies of FQH systems, and an exact solution for
a disorder-free finite-size Hofstadter model \cite{PhysRevB.90.115132}. 

In contrast, arbitrarily weak disorder will break translational invariance
and remove the degeneracy of the GSM. (This is a finite size effect,
as the degeneracy remains in the thermodynamic limit.) The states
in the GSM will remain gapped at all flux values. The calculation
of $C_{m}$ is then defined for the $m$-th energy level. We numerically
find that with weak disorder, one state in the GSM randomly obtains
$C=1$, while the other has $C=0$, depending on the specific disorder
configuration. As the disorder strength is increased, the spectral
gap may shrink until the GSM and the first excited set of states are
close enough to strongly mix. In this case, the many-body Chern number
may mix with the excited state manifold and destroy the FQH state.
(Note that level repulsion will prevent the gap from ever exactly
vanishing.) In order to quantify the tolerance to disorder, we define
the minimum of the many-body gap as 
\begin{equation}
\Delta_{{\rm mb}}=\min_{\bm{\gamma}\in{\rm MBZ}}\Delta_{{\rm mb}}\left(\bm{\gamma}\right).\label{eq:Delta-mb}
\end{equation}
This quantity will be used to study the stability of our scheme in
the next section.

\subsection{Robustness to disorder\label{sub:Robustness-to-disorder}}

In any realistic implementation, disorder is present in various forms.
We now study the influence of spatial disorder in (1) interaction
strength, (2) tunneling energy, (3) flux through each plaquette, and
(4) onsite energy upon the interacting Hofstadter model described
above. For each disorder type we calculate the minimum many-body gap,
and then average over many disorder configurations. This disorder
average many-body gap, $\bar{\Delta}_{{\rm mb}}$, will be used as
a heuristic for the disorder tolerance of the FQH state.

We expect the FQH state to be robust to disorder in the interaction
strength, provided that the average interaction strength is large
compared to the tunneling energy. In this blockaded limit, the two-photon
wavefunction overlap is small on any given site, resulting in only
a small energy shift from interaction disorder. (This argument also
applies to the continuum case where the Laughlin wavefunction has
minimal overlap between photons, and therefore quenches interaction
energy.) This expectation is confirmed through exact diagonalization:
The stability of the many-body gap was calculated by taking the smallest
gap as a full $2\pi$ flux was inserted through the twisted boundary
conditions, as in Eq. \eqref{eq:Delta-mb}. This calculation was performed
for a range of average interaction strengths up to $\bar{U}_{i}\sim20t$,
with a disorder strength $\delta U_{i}=U_{i}-\bar{U}$ uniformly distributed
in the different ranges of $\pm\bar{U}\times\{\frac{1}{4},\frac{1}{2},\frac{3}{4},1\}$.
The results are shown in Fig. \ref{fig:FQH-numerics}(b) demonstrating
that interaction disorder does not significantly reduce the many-body
gap. We further numerically verified that for all disorder configurations
the many-body gap remained open throughout flux insertion, and the
total many-body Chern number in the ground state manifold remained
$C=1$. 

It is curious that even for the case of the strongest interaction
disorder (${\rm max}_{i}|\delta U_{i}|=\bar{U}$) when certain lattice
sites can be almost non-interacting, the topological state survives
with a many-body gap that does not seem to be limited by the smallest
$U_{i}$ in the system. In our system, we still expect only a small
probability for the overlap of two photons, and therefore a robustness
to completely removing some number of qubits. Removing qubits may
provide additional flexibility to the proposed experimental realization
of the interacting lattice, as the ferrites require a large bias DC
magnetic field that is detrimental to the operation of superconducting
qubits. We therefore repeat the previous calculation, now with interactions
turned off entirely on all ferrite sites (see Fig. \ref{fig:FQH-numerics}(c)).
We find the topological ground state is preserved, although the disorder-averaged
many-body gap $\bar{\Delta}_{{\rm mb}}$ is reduced by a factor of
$\ensuremath{\sim4}$.

For interaction disorder we considered a wide distribution which will
realistically arise due to fabrication variations in qubits. By contrast,
tunneling disorder will be small due to a combination of precision
machining techniques and the quadratic insensitivity of the dark state
on ferrite sites. We still consider small to moderate disorder in
both amplitude ($\left|\delta t_{ij}\right|$, uniformly distributed)
and phase of the tunneling rates; the disorder in phase translates
to a (Gaussian distribution) disordered flux in each plaquette of
$\delta\alpha$. The resulting many-body gaps are plotted in Fig.
\ref{fig:FQH-numerics}(d)-(e) when the photon-photon interactions
are assumed to be large. We find the many-body gap decreases only
linearly for both tunneling disorder types in this regime, suggesting
that tunneling disorder will not be a significant issue in experiment. 

The system is more sensitive to onsite disorder which shifts the onsite
mode frequency to $\omega_{0}\rightarrow\omega_{0}+\delta\omega_{i}$.
We explore this physics in Fig. \ref{fig:FQH-numerics}(f) by adding
random onsite (uniformly distributed) disorder in the exact diagonalization
calculation. The situation is similar to previous results for disordered
(fermionic) FQH systems \cite{PhysRevLett.90.256802,PhysRevB.72.075325}:
The total many-body Chern number of the ground state manifold is probabilistically
distributed between the lowest lying states for small disorder. At
larger disorder a collapse of a mobility gap results in a transition
to an insulating state. We find that for weak disorder, the many-body
gap persists, but declines roughly linearly in the disorder strength.
Using 3D microwave cavities, onsite disorder is given by variations
in the cavity resonances determined by the cavity dimensions and can
precisely controlled with modern machining techniques. Additionally,
the onsite resonance can be made tunable, over a range larger than
the tunneling $t$, by e.g. using a tuning screw or piezo stack to
perturb critical dimensions of the cavity \cite{matthaei1964microwave}.
Combined with tomography techniques \cite{1367-2630-13-1-013019,simon-tomography}
which can map out a tight-binding Hamiltonian, onsite disorder can
be characterized in the fully coupled lattice, and further reduced.

\section{Discussion}

We have presented a scalable 3D microwave circuit architecture to
explore bosonic FQH models of photons. Central to this approach is
a method for implementing the necessary TRSB flux of the single particle
band structure: the onsite mode structure is used to induce this flux,
rather than the tunnel couplers themselves. Specifically, a ferrite
is used to couple a degenerate cavity manifold, resulting in an isolated
cavity eigenmode at frequency $\omega_{0}$ that has uniform phase
winding around the cavity periphery. As a photon tunnels in and out,
it will accumulate the phase difference between the input and output
tunnel contact. This phase is a geometric (Berry's) phase, and will
contribute a nontrivial flux to any connected plaquette. This allows
for exploration of, e.g., an $\alpha=1/4$ Hofstadter model on a square
lattice. The circuit elements are entirely passive, providing a distinct
advantage from competing protocols that require driving of the tunnel
matrix elements.

In practice, 10GHz microwave resonators may be realized with ferrite-time
reversal breaking of order $g_{F}=400\,{\rm MHz}$ \cite{Kostylev2016};
this allows a Hofstadter model bandwidth of $200\,{\rm MHz}$ for
tunneling energy of $J=50\,{\rm MHz}$, resonator-to-resonator disorder
of order $\sim1\,{\rm MHz}$, and resonator linewidths of $\kappa\sim20\,{\rm Hz}$
for superconducting coaxial resonators \cite{Reagor2013-highQ}, and
$\kappa\sim10\,{\rm kHz}$ in modest magnetic fields. For standard
transmon qubits, $U=350\,{\rm MHz}$ is readily achieved \cite{PhysRevLett.107.240501},
providing a many-body gap of order (see above) of $\Delta_{{\rm mb}}\sim8\,{\rm MHz}$
, which is easily resolvable. Such states may be spectroscopically
populated \cite{Schausz:2012aa}, and their average occupation of
1/8 per site verified by site-resolved spectroscopy \cite{Houck:2012aa}.

This lattice architecture then forms the basis for the investigation
of topological many-body physics. Effective photon-photon interactions
are implemented through the addition of superconducting qubits. Consistent
with previous works, we predict the emergence of a $\nu=1/2$ bosonic
FQH eigenstate, even at a large flux per plaquette of $\alpha=1/4$.
We further verify that this phase is relatively robust to experimentally
realistic disorder in onsite energy, interactions, tunneling energy,
and flux in a plaquette. In conjunction with state of the art proposals
to implement chemical potentials for photons \cite{hafezi2013non,hafezi2015chemical}
this work provides a complete roadmap to photonic fractional quantum
hall physics, and a path to spectroscopic creation and manipulation
of anyons \cite{paredes20011,umucalilar2013many}.

\section{Acknowledgements}

This work was supported by ARO grant W911NF-15-1-0397; D.S. acknowledges
support from the David and Lucile Packard Foundation; B.M.A. and R.M.
acknowledge support from the UChicago MRSEC grant NSF-DMR-MRSEC 1420709;
C.O. is supported by the NSF GRFP. The authors thank D. Angelakis,
W. DeGottardi, M. Hafezi, and M. Levin, for helpful discussions.

\bibliography{TopCav}

\end{document}